\documentclass[doublecol]{epl2}

\usepackage[colorlinks=true,linkcolor=black, urlcolor = black, citecolor = blue, anchorcolor = blue]{hyperref}

\usepackage{amsfonts, amsmath, amsthm, amssymb} 
\usepackage{graphicx}
\usepackage[tight]{subfigure}
\subfiguretopcaptrue
\usepackage{multirow}
\usepackage{xcolor}
\usepackage{xspace}
\graphicspath{{pix/}}

\newcommand{\etax}[1]{\eta_{x,#1}}

\newcommand{\ave}[1]{\left\langle{#1}\right\rangle}

\newcommand{\latin}[1]{{\it #1}}
\newcommand{\ie}{\latin{i.e.}\@\xspace}
\newcommand{\eg}{\latin{e.g.}\@\xspace}

\newcommand{\plaind}{\text{d}}

\newcommand{\elabel}[1]{\label{eq:#1}}
\newcommand{\eref}[1]{(\ref{eq:#1})}
\newcommand{\Eref}[1]{Eq.~(\ref{eq:#1})}
\newcommand{\Esref}[1]{Eqs.~(\ref{eq:#1})}

\newcommand{\Figref}[1]{Fig.~\ref{fig_#1}}
\newcommand{\bungledXR}[2]{#2}

\newcommand{\wasrevised}[1]{#1}

\begin{document}

\title{Correlations and hyperuniformity in the avalanche size of the Oslo Model}
\shorttitle{Hyperuniformity in the avalanche size of the Oslo Model}

\author{R.~ Garcia-Millan\thanks{ \email{garciamillan16@imperial.ac.uk}} \inst{1,2}
 \and G.~ Pruessner\thanks{ \email{g.pruessner@imperial.ac.uk}} \inst{1,2}
 \and L.~ Pickering \inst{3}
 \and K.~ Christensen \inst{2,3}
 }
\institute{                    
  \inst{1} Department of Mathematics, Imperial College London, London SW7 2AZ, United Kingdom\\
  \inst{2} Centre for Complexity Science, Imperial College London, London SW7 2AZ, United Kingdom\\
  \inst{3} Department of Physics, Blackett Laboratory, Imperial College London, London SW7 2AZ, United Kingdom
}
\shortauthor{R.~ Garcia-Millan \etal}

\newcommand{\XRefereeLabel}[1]{\wasrevised{[#1]}}
\newcommand{\RefereeLabel}[1]{}


\abstract{
Certain random processes display anticorrelations resulting \RefereeLabel{B2} in local Poisson-like disorder and global order, where correlations suppress fluctuations. Such processes are called hyperuniform. Using a map to an interface picture we \RefereeLabel{B5} \wasrevised{show via analytic calculations} that a sequence of avalanche sizes of the Oslo Model is hyperuniform in the temporal domain with the minimal exponent $\lambda=0$. We identify the conserved quantity in the interface picture that gives rise to the hyperuniformity in the avalanche size. We further discuss the fluctuations of the avalanche size in two variants of the Oslo Model. We support our findings with numerical results.
}

\maketitle

\newcommand{\rsection}[1]{\section{#1}}
\newcommand{\subrsection}[1]{\subsection{#1}}

We study the fluctuations of the avalanche size $s$ of the Oslo rice pile Model \cite{ChristensenETAL:1996},  a paradigmatic example of a non-equilibrium system that evolves spontaneously into a scale-invariant state thus considered a representative case of Self-Organised Criticality \cite{ChristensenETAL:1996, PhysRevLett.77.111, pruessner2003oslo, PhysRevE.94.042314, pruessner2012self, corral2004calculation, dhar2004steady}.
Because of the complexity of the model and the number of random variables involved, little is understood about the correlations between avalanches. For example,  when supposed asymptotics of moments are studied, often probability density functions are sampled by deliberately generating effectively independent samples \cite{najafi2012avalanche}. As we show in the present work, correlations in the avalanche size of the Oslo Model \emph{suppress} fluctuations giving rise to a phenomenon that has been dubbed hyperuniformity, briefly outlined below.

The fluctuations in a time series of random variables $s_t$, $t\in\mathbb{N}$, may be characterised by the variance
\begin{equation}
\sigma^2(S(M))\equiv\left\langle S^2(M) \right\rangle - \langle S(M) \rangle ^2 \propto M^\lambda,
\elabel{eq_fluc}
\end{equation}
where $\langle\bullet\rangle$ denotes an expectation, $S$ is the sum $S(M)=s_1+\ldots+s_M$ and $\lambda$ is an exponent to be deternined. The variance of the unbiased  \cite{brandt1999data} estimator $\overline{s}=S(M)/M$ of the mean  $\langle s \rangle$ is $\sigma^2\left( \overline{s}(M) \right) = \sigma^2(S(M))/M^2\propto M^{\lambda-2}$.

If $s_t$ are independent and identically distributed, then $\lambda=1$ \cite{van1992stochastic}, but in general, in the presence of correlations, $\lambda$ is not known \cite{papoulis1965probability,gardiner1986handbook,gikhman2015theory}.
Nevertheless, in some particular stochastic processes  \cite{welinder2007multiscaling} and one-dimensional point patterns \cite{Torquato:2003aa, Hexner:2017aa} it has been proved that anticorrelations suppress fluctuations in such a way that $\lambda\in[0,1)$. Such processes are called hyperuniform \cite{Torquato:2003aa,Hexner:2017aa} or superhomogeneous \cite{PhysRevD.65.083523}, and their fingerprint is the suppression of fluctuations on large  scales, manifesting a regularity that is not apparent on short scales \cite{welinder2007multiscaling,zachary2009hyperuniformity, Torquato:2003aa, Hexner:2017aa, hexner2017enhanced,PhysRevD.65.083523,berthier2011suppressed,froufe2016role}.

In this Letter, we \RefereeLabel{B5} \wasrevised{demonstrate} that the variance $\sigma^2\left(\bar{s}(M)\right)$ of the estimator of the mean avalanche size $\bar{s}$ in the one-dimensional boundary driven Oslo Model decays quadratically in $M$, hence showing that the avalanche size is hyperuniform with exponent $\lambda=0$. \RefereeLabel{B1} \emph{To our knowledge, this is the first time that hyperuniformity in the temporal domain has been identified and proven in a well-known interacting particle system.} \RefereeLabel{B1} \wasrevised{Because of the intermittent nature of the dynamics of the Oslo Model, the temporal anti-correlations are communicated from event to event via its configuration in space, rather than by direct interaction as is expected for purely spatial patterns \cite{jiao2014avian}}. We further show that the temporal correlations reduce the fluctuations in the estimates $\overline{s^n}$ of the moments $\langle s^n\rangle$. Consequently, our results imply that we obtain more precise estimates from a set of correlated avalanches than from a set of independent avalanches.
 We support our analytical findings with numerical results and we further explore two variants of the model, namely external drive uniformly distributed on the lattice and the two-dimensional pile.
 
 In our numerical simulations, we compute the variance $\sigma^2\left(\overline{s^n}(M)\right)$ from a sample of $Q$ independent estimates $\left\{\overline{s_1^n}(M), \ldots, \overline{s_Q^n}(M)\right\}$ of $\ave{s^n}$, where each $\overline{s_q^n}(M)$ is computed from $M$ consecutive avalanches, $\overline{s_q^n}(M) = \left(s_{q,1}^n+\ldots+s_{q,M}^n\right)/M$. All our measurements have been taken once the pile is in a recurrent configuration, that is when its statistical properties are stationary \cite{dhar2004steady,paczuski1996avalanche,corral2004calculation,PhysRevE.78.041102,PhysRevLett.77.111,pruessner2012self}.
 
\RefereeLabel{B5} Our proof of $\sigma^2\left(\bar{s}(M)\right)\propto M^{-2}$ has the following structure: starting from the description of the Oslo Model \cite{ChristensenETAL:1996} in an interface picture \cite{PhysRevLett.77.111, pruessner2003oslo} we are able to write the sum $S(M)$ of the sizes $s_1,\ldots,s_M$ of $M$ consecutive avalanches as the sum \eref{eq_SM} of a deterministic term and a bounded noise asymptotically independent of $M$ \eref{eq_xi1}, so that the variance of the sum converges in large $M$. It follows that the  avalanche size ${s}$ is hyperuniform for large $M$ with $\lambda=0$. 
\RefereeLabel{B5} Although our proof of hyperuniformity does not draw on scaling arguments,
for the sake of completeness, we study the scaling form of  $\sigma^2\left
(\bar{s}(M)\right)$ in the entire domain of $M$ and show that it displays
a crossover
from Poisson-like behaviour (\latin{i.e.}~$\lambda=1$) to hyperuniformity (with $\lambda=0$).
For simplicity, our argument is applied to $d=1$ although it can be generalised to higher dimensions.
 
\rsection{The Oslo Model}
\label{sect_oslo}
In the Oslo Model \cite{ChristensenETAL:1996} $n_x$ particles reside on site $x\in\{1,\ldots,L\}$ of a one-dimensional lattice of size $L$ and its configuration is described by the set of local slopes $z_x=n_x-n_{x+1}$ with $n_{L+1} \equiv 0$. Each site has associated with it a randomly chosen critical slope $z_x^c\in\{1,2\}$. Given $z_x\leq z_x^c$ for all $x$, which is called a stable or quiescent configuration, the evolution of the pile follows the steps (i) \emph{drive}, a grain is dropped at site $x=1$ so that $z_1\to z_1+1$; (ii) \emph{relaxation}, every unstable site $x$, i.e. with $z_x>z_x^c$, is relaxed in parallel with all other unstable sites by toppling one grain to the neighbouring site on its right, resulting in the update rules $z_x\to z_x-2$ and $z_{x\pm1}\to z_{x\pm1}+1$ in the bulk. The update rules that apply to the boundaries are $z_1\to z_1-2$, $z_2\to z_2+1$ (open) and $z_L\to z_L-1$, $z_{L-1}\to z_{L-1}+1$ (closed). After every toppling, $z_x^c$ is redrawn to be either 1 or 2 with equal probability. The relaxation process is repeated until the entire pile is in a stable configuration. The totality of the relaxations constitute an avalanche. We distinguish between the macroscopic timescale denoted by integer $t$, where the pile evolves from one stable configuration to another as $t$ increases by $1$, and a much smaller microscopic timescale in which the avalanches unfold.

The avalanche size $s$ is defined as the total number of topplings that occur between two consecutive quiescent states. We find numerically that the temporal correlations between avalanches $\text{Cov}(s_i,s_{j})=\ave{s_is_{j}}-\ave{s_i}\ave{s_{j}}$ are negative for $i\neq j$ and decay approximately exponentially (see \bungledXR{}{Sec.~S1}), with a correlation time that is a power law in $L$. In that sense, $s_i$ has short-ranged correlations. As we will show in the following, the anticorrelations exactly cancel the first term of order $1/M$ in the variance $\sigma^2\left(\bar{s}(M)\right) = \sigma^2\left(s\right)/M + \sum_{i,j=t+1; i\neq j}^{t+M}\text{Cov}(s_i,s_j)/M^2$ so that $\sigma^2\left(\bar{s}(M)\right) \sim M^{-2}$ in large $M$.

\rsection{The interface picture}
\label{sect_interf}
\begin{figure}[t]
\centering
 \includegraphics[width=0.49\textwidth]{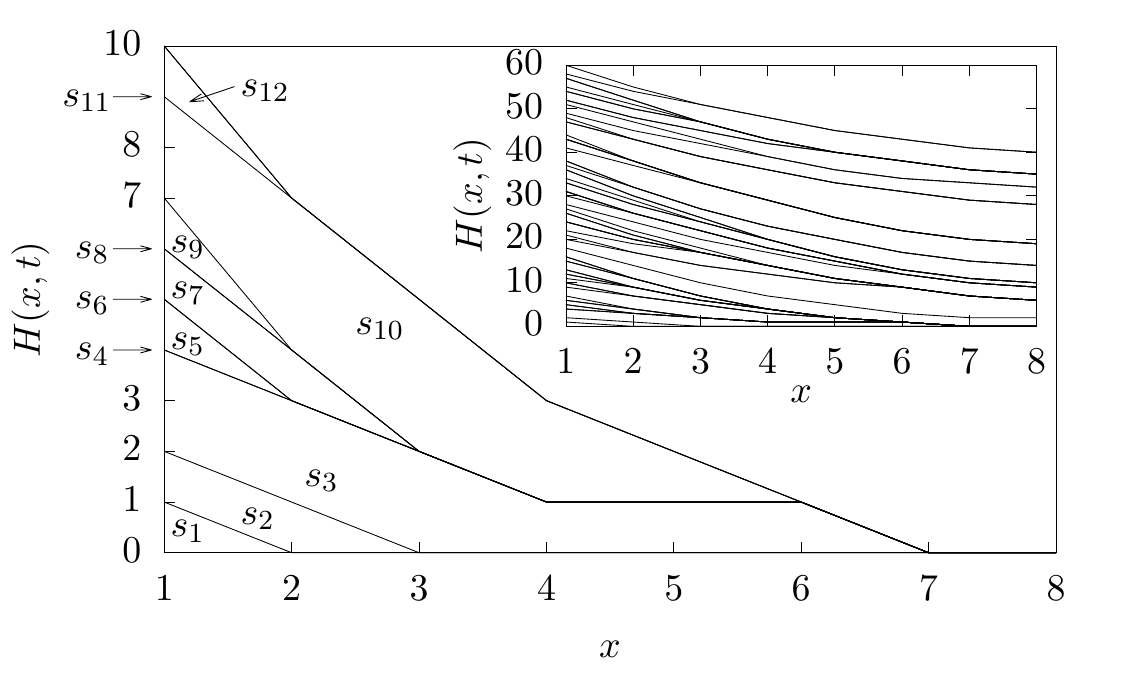}
  \caption{ \label{fig_s_sketch}
 Example of the interface $H(x,t)$ of a one-dimensional boundary driven pile of size $L=8$ with initial state $z_x=1$ and $z_x^c$ initially chosen at random for all $x$. In this realisation of the process, the series of avalanche sizes is $\{1,2,9,0,1,0,2,0,1,11,0,1,\ldots\}$, consistent with \Eref{eq_s_H}. The avalanche sizes are exactly the area between consecutive interface configurations if the interfaces are drawn using bars. Inset: $t\in[0,65]$.}
 \end{figure}

The dynamics of the pile can equivalently be described by the function $H(x,t)$, which gives the total number of topplings at site $x$ up to time $t$ \cite{PhysRevLett.77.111, pruessner2003oslo}. In this picture, the avalanche size $s$ is equal to the area enclosed between two consecutive stable interface configurations,
\begin{equation}
s_t=\sum_{x=1}^L\big(H(x,t+1)-H(x,t)\big),
\elabel{eq_s_H}
\end{equation}
 as illustrated in \Figref{s_sketch}.
 
The temporal evolution of $H(x,t)$ was set out in \cite{pruessner2003oslo} for one particular choice of the initial condition of the pile and two open boundaries. Here we use a generalised version for an arbitrary stable initial condition and one open and one closed boundary, see \bungledXR{\ref{sect_app_interface}}{Sec.~S2}. In this setup, the spatial fluctuations of the interface $H(x,t)$ are confined because its curvature is bounded, \bungledXR{}{Eq.~(S9)}.
  
We define the multiple drive avalanche size $S$ as the sum $S(t,M)=\sum_{i=t+1}^{t+{M}} s_i $, 
based on $M$ consecutive avalanche sizes $s_i$,
and the estimator $\bar{s}$ of $\langle s \rangle$ as $\bar{s} = S(t,M)/M$. By the Abelianess of the Oslo Model \cite{dhar2004steady}, the total avalanche size triggered by $M$ simultaneous charges is identical to the sum of $M$ consecutive avalanche sizes. Since the macroscopic timescale is measured in units of the number of grains added to the pile, $M$ is simultaneously a measure of time, drive size and sample size of $\bar{s}$.

\begin{figure}
 \includegraphics[width=0.49\textwidth]{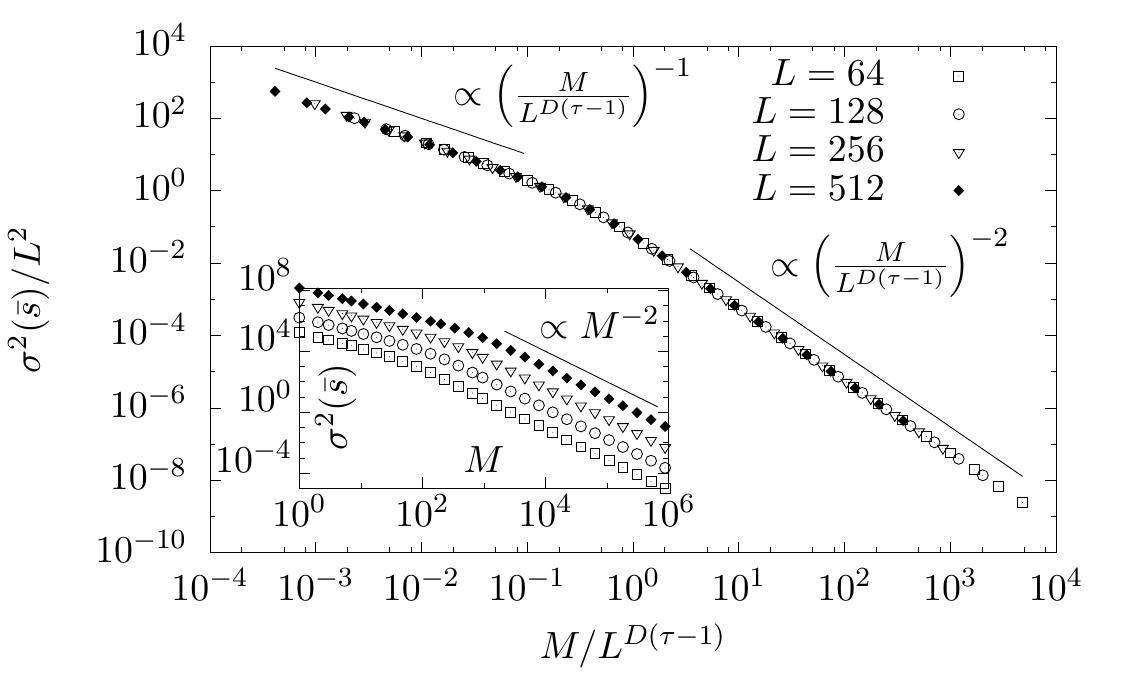}
 \caption{ \label{fig_Dx} Data collapse of $\sigma^2\left(\bar{s}\right)$ \wasrevised{across many orders of magnitude} as a function of $M/L^{D(\tau-1)}$ according to \Eref{eq_Dx}. The variance is compared to plain power laws, shown as straight lines, in the two regimes $M\ll L^{D(\tau-1)}$ and  $M\gg L^{D(\tau-1)}$ with $M\in\left[1,10^6\right]$. Inset: unscaled $\sigma^2\left(\bar{s}\right)$ as a function of $M$.} 
 \end{figure}
 
 \begin{figure}
 \includegraphics[width=0.49\textwidth]{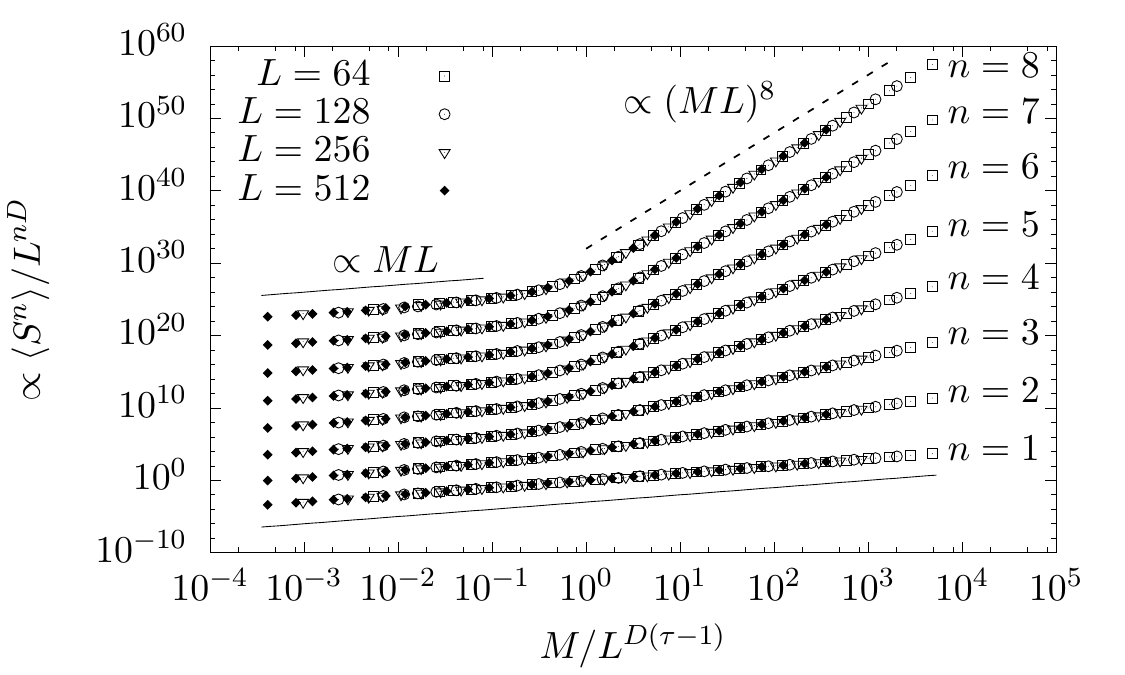}
\caption{ \label{fig_Sn1} Data collapse of the moments $\left\langle S^n(M;L)\right\rangle$ \wasrevised{across many orders of magnitude} according to \Eref{eq_Sn} with $n\in\{1,\ldots,8\}$ as a function of $M/L^{D(\tau-1)}$ for the system sizes $L\in\{64,128, 256, 512\}$ and $M\in\left[1,10^6\right]$. \RefereeLabel{B6} \wasrevised{The collapsed curves for $n\geq2$ have been shifted vertically for better visibility.} The full line, with slope $1$, is shown to emphasise the linearity of $\left\langle S^n\right\rangle$ in small $M$; the dashed line, with slope $8$ shows the proportionality $\left\langle S^8\right\rangle\propto(ML)^8$ in large $M$.}
\end{figure}

Using \Eref{eq_s_H}, the multiple drive avalanche size $S$ is equal to the area enclosed by the interfaces $H(x,t+M)$ and $H(x,t)$,
 \begin{equation}
 S(t,M)  = \sum_{x=1}^L \big(H(x,t + {M}) - H(x,t)\big).
 \elabel{eq_def_S}
 \end{equation}
The function $H(x,t)$ can be determined in closed form in terms of the random variable $\etax{t}\in\{-1,0,1\}$ \cite{pruessner2003oslo}, which encodes the configuration of the pile via the relation $\etax{t}=1-z_x$ at each time $t$, see \bungledXR{\ref{sect_app_interface}}{Sec.~S2}.
We find that $S(t,M)$ can be expressed in closed form as the sum of a deterministic term and a noise
 \begin{equation}
  S(t,M) = ML + \xi(t,M),
   \elabel{eq_SM}
 \end{equation}
 where $\xi(t,M)$ can be written in the form \Eref{eq_xi1},
\begin{eqnarray}
\xi(t,M) &=& \frac{1}{2}\sum_{x=1}^{L}\bigg[\eta_{x,t+{M}} -\eta_{x,t} \\
&&+ \left(x(L+1-x)+\frac{1}{2}x(x-1)\right)\nonumber\\
&&\times\bigg(\eta_{x-1,t+{M}}+ \eta_{x+1,t+{M}} - \eta_{x-1,t}-\eta_{x+1,t}\bigg)\bigg]\nonumber
\elabel{eq_xi1}
\end{eqnarray}
which is, in fact, a finite sum of bounded random variables. \RefereeLabel{B4} \wasrevised{The strict bounds of $\xi$ are $\pm\left[\frac{2}{3}L(L+1)(L+2)-L^2\right]\in\mathbb{Z}$ and, asymptotically, it is easy to see that the distribution of $\xi$ neither depends on $M$ nor on $t$.} 
In other words, the values that $\xi$ can take do not scale with $M$ \RefereeLabel{B4} \wasrevised{for large $M$}. From \Eref{eq_xi1} we deduce that  $\langle \xi\rangle =0$, as it is the difference of two stationary expectations, and therefore $\langle S(t,M)\rangle=ML$. \wasrevised{Moreover,} \Eref{eq_SM} further implies the estimate of the mean is $\bar{s}(M)=L+\xi/M$. \RefereeLabel{B5} Crucially, $\ave{\xi^2}$ does not scale in $M$, so that the variance of $\bar{s}$, \wasrevised{which is} $\sigma^2\left(\bar{s}(M)\right)= \left\langle\xi^2\right\rangle/M^{2}$, decays quadratically in the limit of large $M$, proving that \emph{$s$ is hyperuniform in this limit} (see inset of \Figref{Dx}) with exponent $\lambda=0$,  the fingerprint of enhanced hyperuniformity \cite{hexner2017enhanced}.

\subrsection{Scaling of $\sigma^2\left(\bar{s}\right)$}
In the following, we derive the scaling of the moments $\ave{S^n}$ of the
total avalanche size $S$ resulting from $M$ drives and show that the hyperuniformity
of $s$ is consistent with its established scaling.

The single drive avalanche size $s$, which is $S$ with $M=1$, is power-law distributed \cite{ChristensenETAL:1996, PhysRevE.94.042314,paczuski1996avalanche, pruessner2012self} and its moments are $\ave{s^n}\sim L^{D(1+n-\tau)}$, where $D=2.250\pm0.002$ is the avalanche dimension and $\tau=1.5556\pm0.0005$ \cite{ChristensenETAL:1996,
PhysRevE.94.042314,paczuski1996avalanche} is the avalanche size exponent. The characteristic avalanche size is $L^D$. Since $\ave{s}= L$, the critical exponents satisfy $D(2-\tau)=1$.

Another characteristic scale entering the scaling of $\ave{S^n}$, and the $n$-point correlation function of $H$ below, is the drive size $M^*$ after which two interface configurations are effectively independent. The interfaces $H(x,t_1)$ and $H(x,t_2)$ can be considered independent if they are completely detached, \ie $H(x,t_2)>H(x,t_1)$ for all $x$. This is typically the case when their separation exceeds the characteristic vertical fluctuations of the interface, which scale with $L^{\chi}$ \cite{PhysRevLett.77.111,pruessner2012self}, so that $M^* \propto L^{\chi}$, where $\chi$ is the roughness exponent. Using $D=\chi+d$ \cite{paczuski1996avalanche, PhysRevLett.77.111} and $D(2-\tau)=d$ in the case of boundary driving, we have $\chi=D(\tau-1)$.

In Sec.~S3 we discuss why the scale $M^*$ is a necessary argument in the $n$-point correlation function of $H$. Assuming that the $n$-point correlation of $H$ displays standard gap-scaling, \eg $\langle H(x_1,t_1)\ldots H(x_n,t_n)\rangle\propto L^{n\chi}
\mathcal{C}_n\left(\ldots,|t_2-t_1|/aL^{D(\tau-1)},\ldots\right)$, \bungledXR{\Eref{eq_corr_func2}}{Eq.~(S17)}, where the scale $L^{D(\tau-1)}$ enters explicitly and $a$ is a metric factor, it follows that
\begin{eqnarray}
\ave{S^n} &=& \left\langle\int\ldots\int \big( H(x_1,t+M)-H(x_1,t)\big)\ldots\right.
\elabel{eq_Sn}\\
&&\bigg.\ldots \big( H(x_n,t+M)-H(x_n,t)\big) \plaind^dx_1\ldots\plaind^dx_n \bigg\rangle \nonumber\\
&=& L^{nd+n\chi}\mathcal{G}_n \left(\frac{M}{L^{D(\tau-1)}}\right) = L^{nD}\mathcal{G}_n \left(\frac{M}{L^{D(\tau-1)}}\right),
\nonumber
\end{eqnarray}
with $\mathcal{G}_n$ a scaling function. The scaling of $\ave{S^n}$ is compatible with the case $M=1$, $\ave{s^n}\sim L^{D(1+n-\tau)}$, only if the scaling functions $\mathcal{G}_n$ are linear for small arguments (see \Figref{Sn1}), so that $\ave{S^n}\propto M L^{D(1+n-\tau)}$ in small $M$ for all $n>\tau-1$. For large arguments, $\ave{S^n}\propto (ML)^n$ from \Eref{eq_SM}, as $\xi$ does not scale in $M$.

Writing $\mathcal{D}(x) = x^{-2}\left(\mathcal{G}_2(x)-\mathcal{G}_1^2(x)\right)$, the scaling form of $\sigma^2\left(\bar{s}\right)$ is thus
\begin{equation}
\sigma^2\left(\bar{s}\right) = \frac{\sigma^2\left(S(t,M)\right)}{M^2} =L^2 \mathcal{D}\left(\frac{M}{L^{D(\tau-1)}}\right),
\elabel{eq_Dx}
\end{equation}
where $\mathcal{D}$ is a scaling function whose asymptotes we characterise in the following (see \Figref{Dx}). Firstly, $\mathcal{D}(x)\sim x^{-1}$ for $x\ll 1$, as $\mathcal{G}_n(x)\propto x$ for small $x$. For large $M\gg M^*$, on the other hand, the boundedness of $\xi=S-ML$, \Eref{eq_SM}, implies finite $\sigma^2(S)$ and thus hyperuniformity of $s$ with exponent $\lambda=0$, so $\mathcal{D}(x)\sim x^{-2}$ for $x\gg 1$. The crossover of the scaling takes place at $M^*\propto L^{D(\tau-1)}$. This concludes our derivation of the scaling of the variance $\sigma^2\left(\bar{s}\right)$.

\begin{figure}
 \includegraphics[width=0.49\textwidth]{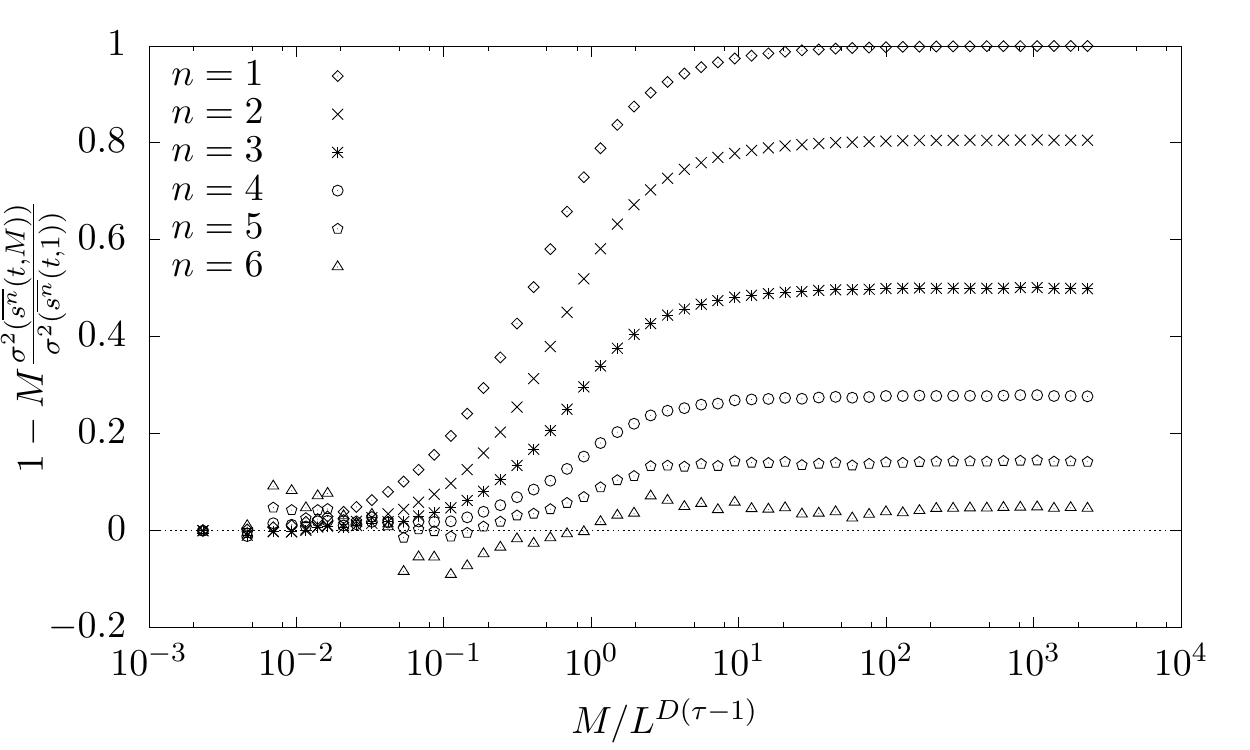}
 \caption{ \label{fig_hyp_sn_0} Quantity $1-M\left(\sigma^2\left(\overline{s^n}(t,M)\right)\right)/\left(\sigma^2\left(\overline{s^n}(t,1)\right)\right)$ that compares $\sigma^2\left(\overline{s^n}(t,M)\right)$ with $\sigma^2\left(\overline{s^n}(t,1)\right)/M$, which corresponds to the variance of the estimate $\overline{s^n}$ calculated from $M$ independent samples, for the system size $L=128$ and $M\in\left[1,10^6\right]$.}
 \end{figure}

\rsection{Higher moments}
In the following, we investigate the fluctuations of higher moments.
The estimator of $\langle s ^n\rangle$ is $\overline{s^n}(t,M)=\left(\sum_{i=t+1}^{t+M}s_i^n\right)/M$. \Figref{hyp_sn_0} shows the quantity $1-M\left(\sigma^2\left(\overline{s^n}(t,M)\right)\right)/\left(\sigma^2\left(\overline{s^n}(t,1)\right)\right)$, which tends to $1$ if $s^n$ is hyperuniform, lies in $(0,1)$ if there is suppression of fluctuations (in amplitude but not in rate of convergence) and tends to $0$ if it shows Poisson-like behaviour.
Although the random variables $s^n$ with $n>1$ are not hyperuniform, our results show that the correlations among consecutive avalanches still suppress fluctuations even in higher moments.
 
\rsection{Discussion}
Our results show that the avalanche size of the Oslo Model has the typical behaviour of hyperuniform systems \cite{Torquato:2003aa,Hexner:2017aa,zachary2009hyperuniformity,berthier2011suppressed}: apparent Poison-like disorder on short scales and suppression of fluctuations on large scales. The suppression of fluctuations is due to the negative correlations between consecutive avalanches that die off on the scale of $M^*\propto L^{D(\tau-1)}$, which is precisely proportional to the correlation time. \RefereeLabel{A2} We have identified the \RefereeLabel{A2} \wasrevised{key ingredients} of hyperuniformity in \RefereeLabel{A2} \wasrevised{the Oslo Model on the basis of} the interface picture (\Figref{s_sketch}), namely the \emph{conservation} of the area $ML$ between interfaces, which is due to (a) the boundedness of the curvature of the interfaces and (b) the drive fixed on one site.

The boundedness of the curvature of the interfaces is explained by the stochastic dynamics that governs $H(x,t)$ (\bungledXR{\ref{sect_app_interface}}{Sec.~S2}, in particular \bungledXR{}{Eq.~(S9)}) which encodes the distribution rule of slope units during the relaxation ($z_x\to z_x-2$ and $z_{x\pm1}\to z_{x\pm1}+1$). Further analysis shows that hyperuniformity is found provided only that the driving site is fixed and that at least one boundary is open, which is a necessary condition for stationarity. 
Regarding the numerical aspects of the Oslo Model, the rapid convergence of $\bar{s}$ means that for a fixed sample size $M$, more precise statistics are obtained when the samples are taken sequentially rather than independently.

One may ask to what extent hyperuniformity is a particular feature of the Oslo Model or a generic property of Self-Organised Criticality. To this end, we have studied the Oslo Model in two other settings\wasrevised{,} \RefereeLabel{A1} \wasrevised{namely the bulk-driven Oslo Model and the two-dimensional Oslo Model}.

Firstly, we have derived the multiple drive avalanche size $S\wasrevised{(M)}$ for the bulk-driven Oslo Model \cite{PhysRevE.78.041102, pruessner2012self} in closed form\wasrevised{, see} \bungledXR{\ref{sect_app_bulk}}{Sec.~S4}.
  We find that the fluctuations due to the random drive (uniformly distributed in the bulk), whose variance can be calculated exactly, prevent hyperuniformity in the avalanche size. \RefereeLabel{A1} Because of this driving, the curvature of the interfaces is not bounded, so the area between them is no longer conserved.
  Asymptotically in large $M$, the variance of $\bar{s}$ is $\sigma(\bar{s})\propto L^4M^{-1}$; numerically we find its scaling form is
$\sigma^2(\bar{s}) = L^{3.5}\tilde{\mathcal{D}}\left({M}/{L^{0.5}}\right)$,
with $\tilde{\mathcal{D}}(x)\propto x^{-1}$ as $x\to\infty$ (see Tab.~\ref{tab1} and \bungledXR{\Figref{oobulk}}{Fig.~S3}).
  
 Secondly, we have studied \wasrevised{numerically} the two-dimensional Oslo Model with fixed driving position (corner) and with uniformly distributed driving (bulk drive), see Tab.~\ref{tab1} and \bungledXR{\ref{sect_app_2d}}{Sec.~S5}. In the former case, we recover hyperuniformity for the same reasons as in 1D and in the latter we find Poisson-like behaviour for the same reasons as in 1D.

\begin{table}[h]
\caption{\label{tab1} Summary of our results for each of the settings studied: the exponent $\sigma_1$ of $\ave{s}\sim L^{\sigma_1}$, the crossover point $M^*$ of $\sigma^2(\bar{s})$ and its scaling behaviour for large $M$.}
\begin{tabular}{| c @{\hspace{1pt}} c @{\hspace{1pt}} c @{\hspace{4pt}}c @{\hspace{1pt}} l |}
\hline
$d$	&	Drive	& $\sigma_1$ & $M^*$ & Large $M$ behaviour\\
\hline
\multirow{2}{*}{1} & corner & $1$ & $L^{D(\tau-1)}=L^\chi$ & \small{hyperuniform,} $\sim M^{-2}$\\
		 		& \small{uniform bulk} & $2$ & $L^{0.5}$ & \small{Poisson-like,} $\sim M^{-1}$\\
\hline
\multirow{2}{*}{2} & corner & $0$ & $L^{D(\tau-1)}=L^{\chi+2}$ & \small{hyperuniform,} $\sim M^{-2}$\\
		 		& \small{uniform bulk} & $2$ & $L^{1.5}$ & \small{Poisson-like, $\sim M^{-1}$}\\
\hline
\end{tabular}
\end{table}

\RefereeLabel{A1} \wasrevised{These results, summarised in Tab.~\ref{tab1}, answer the question of whether hyperuniformity is a particular or a generic feature. We have found two instances of the Oslo Model, the 1D and the 2D Oslo Model with bulk drive, that do not display hyperuniformity. In these cases, the Poisson-like fluctuations introduced by the randomness of the drive dominate over the suppression of fluctuations in the avalanches. The mechanism that causes the anomalous suppression of fluctuations is thus not present in all instances of Self-Organised Criticality.}

In conclusion, we have proven hyperuniformity in the time series of avalanche sizes of the boundary-driven Oslo Model. \RefereeLabel{A1} Its presence is a specific feature of the dynamics of this model. Identifying hyperuniformity in other models, such as the Manna Model, will elucidate its role in interacting particle systems and advance our understanding of Self-Organised Criticality.

\begin{acknowledgments}
We would like to thank Nanxin Wei, Saoirse Amarteifio and Deepak Dhar for helpful discussions, and Andy Thomas for invaluable computing support.
\end{acknowledgments}

\begin{widetext}
\begin{center}
\textbf{\large 
Supplementary Material for  Correlations and hyperuniformity\\ in the avalanche size of the Oslo Model}
\end{center}
\textbf{Abstract} -- In this Supplementary Material we complement the main text with: (Sec.~S1) numerical results for the temporal correlations
between avalanches on the one-dimensional boundary-driven Oslo Model, (Sec.~S2) the map between the  one-dimensional
boundary-driven Oslo Model and the interface picture, (Sec.~S3) scaling form of the higher correlation functions of
the interface, (Sec.~S4) the map between the one-dimensional bulk-driven Oslo Model and the interface picture, and (Sec.~S5)
numerical results on the two-dimensional Oslo Model with boundary (corner) drive and with bulk drive.

\end{widetext}
\shorttitle{SM for Correlations and hyperuniformity in the avalanche size of the Oslo Model}

\pagebreak

\newcommand{\CCbar}{\overline{\mathcal{C}}}
\newcommand{\dint}[1]{\mathchoice{\!\plaind#1\,}{\!\plaind#1\,}{\!\plaind#1\,}{\!\plaind#1\,}}
\newcommand{\cf}{\latin{cf.}\@\xspace}

\setcounter{equation}{0}
\setcounter{figure}{0}
\setcounter{table}{0}
\setcounter{page}{1}
\renewcommand{\theequation}{S\arabic{equation}}
\renewcommand{\thefigure}{S\arabic{figure}}
\renewcommand{\thesubfigure}{(\alph{subfigure})}



\section{S1. Temporal correlations between avalanches}

\begin{figure}[h]
 \centering
\includegraphics[width=0.49\textwidth]{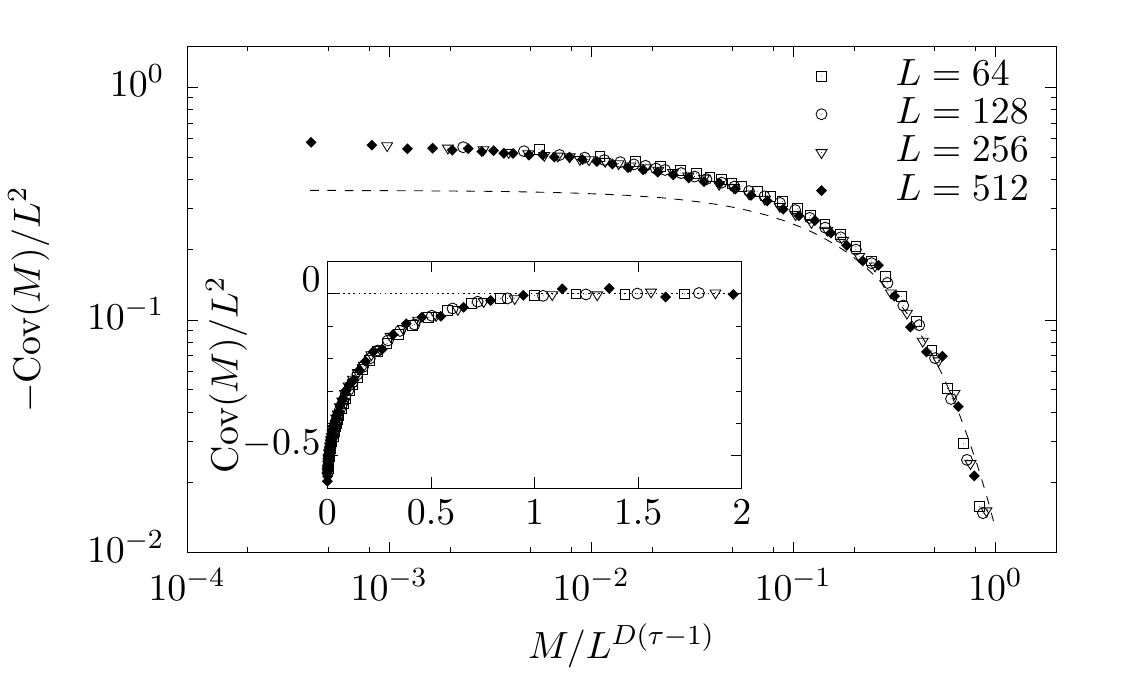}
\caption{ \label{fig_corr} Log-log plot of the rescaled correlation function $\text{Cov}(M)=\ave{s_ts_{t+M}}-\ave{s_t}\ave{s_{t+M}}$ \Eref{eq_corr} obtained numerically for $M>0$, where $s_t$ is taken when the pile is in the stationary regime. The correlations decay exponentially, the dashed line is the function $0.36e^{-x/0.3}$ with $x=M/L^{D(\tau-1)}$. Inset: same data but shown with linear axes, showing that the correlations are negative.}
\end{figure}

Our numerical results, \Figref{corr}, show that the temporal correlations between consecutive avalanches scale in 
\begin{equation}
\text{Cov}(M)=\ave{s_ts_{t+M}}-\ave{s_t}\ave{s_{t+M}}=-L^2\mathcal{F}\left(\frac{M}{L^{D(\tau-1)}}\right),
\elabel{eq_corr}
\end{equation}
which we have fitted against $\mathcal{F}(x)=Ax^{-B}e^{-x/C}$. We find that the exponent $B$ is very small. It decreases with $L$, taking values $B=0.083$ at $L=64$ and $B=0.059$ at $L=512$, which suggests that it may vanish in large $L$ and the $L$-dependent deviation of $\mathcal{F}$ from a pure exponential may thus be regarded as a logarithmic correction. We therefore approximate that $\mathcal{F}(x)=Ae^{-x/C}$ with $A\simeq 0.36$ and $C\simeq 0.30$. Therefore, the correlation time is $CL^{D(\tau-1)}$, which is further discussed in this Letter and in Sec.~S3.

To derive \Eref{eq_corr} we consider the correlation function
\begin{equation}
c(x,x',t,t') = \ave{H(x,t)H(x't')} - \ave{H(x,t)}\ave{H(x't')},
\end{equation}
whose scaling form follows \Eref{eq_2pointH_corr_func_right} discussed further below. From \bungledXR{}{Eq.~(3)} the correlation function between avalanches is calculated as in \Esref{eq_Cov1} to \eref{eq_Cov4} using that $c(x,x',t,t')$ is uniform in $t$ and defining $L^{2D}\mathcal{D} \left(|t-t'|/L^{D(\tau-1)}\right) = \iint c(x,x',t,t')\plaind x\plaind x'$ from \Eref{eq_2pointH_corr_func_right}.
\begin{widetext}
\begin{eqnarray}
\text{Cov}(M)&=&\ave{s_ts_{t+M}}-\ave{s_t}\ave{s_{t+M}} \elabel{eq_Cov1}\\
&=&\ave{\iint\big(H(x,t+1)-H(x,t)\big)\big(H(x',t+M+1)-H(x',t+M)\big)\plaind x\plaind x'} \nonumber\\
&& - \ave{\int\big(H(x,t+1)-H(x,t)\big)\plaind x}\ave{\int\big(H(x',t+M+1)-H(x',t+M)\big)\plaind x'}\elabel{eq_Cov2}\\
&=& \iint \big[2c(x,x',t,t+M) -c(x,x',t+1,t+M) - c(x,x',t,t+M+1) \big]\plaind x\plaind x' \elabel{eq_Cov3}\\
&=& aL^{2D} \left[ 2\mathcal{D} \left(\frac{M}{L^{D(\tau-1)}}\right) -\mathcal{D} \left(\frac{M-1}{L^{D(\tau-1)}}\right) -\mathcal{D} \left(\frac{M+1}{L^{D(\tau-1)}}\right) \right] \elabel{eq_Cov4}
\end{eqnarray}
\end{widetext}
Now, using $\varepsilon = 1/ L^{D(\tau-1)}$ and expanding about $x=M/ L^{D(\tau-1)}$ we have
\begin{equation}
2\mathcal{D}(x)-\mathcal{D}(x-\varepsilon)-\mathcal{D}(x+\varepsilon)=-\varepsilon^2\mathcal{D}''(x) + \mathcal{O}(\varepsilon^4).
\end{equation}
Using the relation $D(2-\tau)=1$, we obtain \Eref{eq_corr} with $a\mathcal{D}(x)=\mathcal{F}(x)$.

\section{S2. Interface picture}
\label{sect_app_interface}

\begin{figure}
 \centering
\subfigure[\hspace{230pt}]{ \label{fig_Ha} \includegraphics[width=0.48\textwidth]{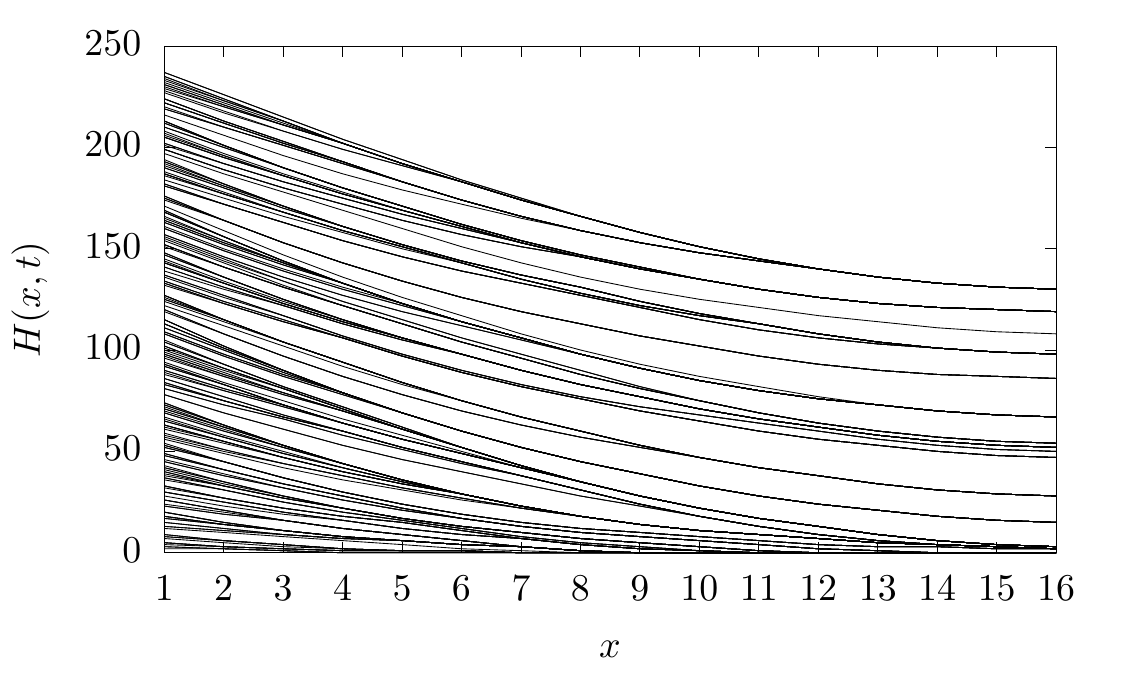}}
\subfigure[\hspace{230pt}]{ \label{fig_Hb} \includegraphics[width=0.48\textwidth]{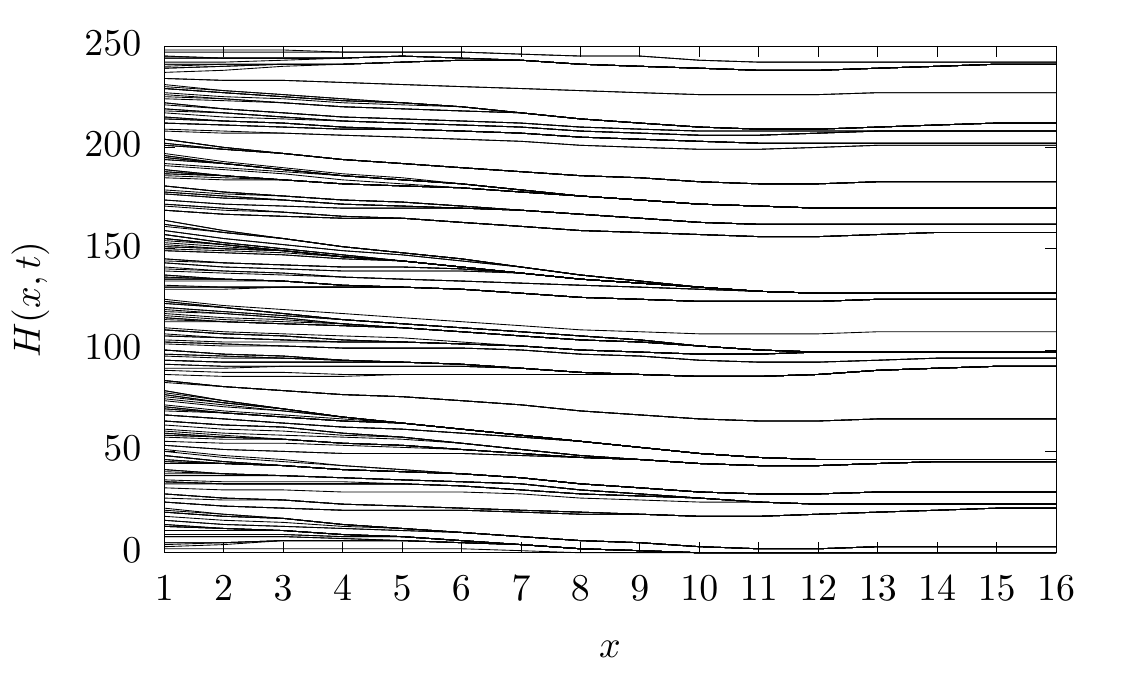}}
\caption{ \label{fig_H} Examples of the interfaces $H(x,t)$ on a one-dimensional boundary-driven pile of size $L=16$ with $t\in[0,250]$. In \subref{fig_Ha} the initial configuration is $z_x=1$ and random $z_x^c$ for all $x$ and in \subref{fig_Hb} the initial state is a recurrent configuration taken at random (more specifically, the configuration of the pile after driving $1000$ times the pile with  $z_x=1$ and random initial $z_x^c$).}
\end{figure}

In this section we demonstrate how the Oslo Model is mapped to an interface depinning model in a generalisation of \cite{pruessner2003oslo,PhysRevLett.77.111}, and solve the stochastic equation of motion to obtain the interfaces $H(x,t)$ (\bungledXR{\Figref{s_sketch}}{Fig.~1} and \Figref{H}), which is needed to derive the exact form of the multiple drive avalanche size as given in \bungledXR{\Esref{eq_SM}}{Eqs.~(4)} and \bungledXR{\eref{eq_xi1}}{(5)}.

 Let $h(x,t)$ count the number of slope units that fall on site $x$ up to time $t$. The function $h(x,t)$ thus counts the number of times the slope $z_x$ is incremented by $1$ as the model evolves, as opposed to $H(x,t)$, which counts the number of times  site $x$ has toppled.
  Let $\eta(x,h(x,t))$ be a quenched noise that takes values $0$ if $h(x,t)$ is even, and $1$ or $-1$ with equal probability if $h(x,t)$ is odd \cite{pruessner2003oslo}.
Although $\eta$ is therefore not explicitly a function of $t$, to ease notation
we write $\eta_{x,t}\equiv \eta(x,h(x,t))$, as for each $x$ and $t$ there
is a unique value of $h(x,t)$. The configuration of the pile in terms of local
slopes $z_x$ (we drop its time dependence solely to ease notation) can be
read from $\eta$ through the relation $\etax{t}=1-z_x$. \emph{Given that $\{z_x\}$ is Markovian by definition of the model, it is clear that $\{\eta_{x,t}\}$ is Markovian as well.}

For a given arbitrary initial configuration of the pile $\{z_x\}$, equivalently $\{\etax{0}\}$, the microscopic temporal evolution of $h(x,t)$ in a microscopic time step of length $\delta t\ll1$ is \cite{pruessner2003oslo}
\begin{eqnarray}
&&\hspace{-10pt}{h}({x,t+\delta t})-{h}({x,t}) =  \frac{1}{2}\Big[{h}({x-1,t}) - 2{h}({x,t})\Big. \nonumber\\
&& \hspace{-10pt}\Big. +{h}({x+1,t})+ \eta_{x-1,t} + \eta_{x+1,t}-\eta_{x-1,0} - \eta_{x+1,0}\Big].\qquad \elabel{eq_QEW1}
\end{eqnarray}
 The boundary conditions that reproduce the original Oslo Model \cite{ChristensenETAL:1996} are: open on the left with an external source $h(0,t)=2 \left\lfloor {t} \right\rfloor\in\mathbb{N}$ that accounts for the drive; and closed on the right $h(L,t)=h(L+1,t)$, giving the boundary conditions of $\eta_{x,t}$, namely $\eta_{0,t}=0$ and $\eta_{L,t}=\eta_{L+1,t}$ for all $t$.

When the pile is in a quiescent state, the left-hand side of \Eref{eq_QEW1} vanishes for all $x$, which gives the expression that governs the function $h(x,t)|_{t\in\mathbb{N}}$ we are interested in,
\begin{equation}
\Delta h(x,t)|_{t\in\mathbb{N}} = a_{x,0}- a_{x,t},
\elabel{eq_gov_h}
\end{equation}
where $\Delta h(x,t)$ denotes the discrete Laplacian and $a_{x,t} = \eta_
{x-1,t}+\eta_{x+1,t}$. As $a_{x,t}$ is bounded, \Eref{eq_gov_h} implies that
$h(x,t)$ has finite curvature, which is, of course, consistent with our numerical
simulations, where the spatial average of $\eta$ in the recurrent state at quiescence is finite and negative.
 The solution of \Eref{eq_gov_h} with the boundary conditions as stated above is
 \begin{equation}
 {h}(x,t)|_{t\in\mathbb{N}} = 2 {t}  - \tilde{h}(x,0) +  \sum_{y=1}^x y a_{y,t} + x \sum_{y=x+1}^La_{y,t},
 \elabel{eq_sol_ht1}
 \end{equation}
  where $ \tilde{h}(x,0) =  \sum_{y=1}^x y a_{y,0} + x \sum_{y=x+1}^La_{y,0}$ is the shift in $h(x,t)$ that accounts for the initial state of the pile. If the initial condition is $\{z_x=1\}$ for all $x$, then $\{\eta_{x,0}=0\}$ and hence $\tilde{h}(x,0) =0$, which is the case studied in \cite{pruessner2003oslo} (see Fig.~S\ref{fig_Ha}).

In \cite{pruessner2003oslo} it is shown that $h(x,t)$ and $H(x,t)$ are related by
\begin{equation}
H(x,t)=\frac{1}{2}\big(h(x,t)+\eta_{x,t}\big).
\elabel{eq_Hh}
\end{equation}
The boundary conditions of $h$ and $\eta$ imply correspondingly $H(0,t)=\left\lfloor {t} \right\rfloor$ and $H(L,t)=H(L+1,t)$ for all $t$ \cite{PhysRevLett.77.111} (see \Figref{H}). 

 Using \Esref{eq_sol_ht1} and \eref{eq_Hh} we have
 \begin{eqnarray}
&&\hspace{-10pt}{H}(x,t)|_{t\in\mathbb{N}} \nonumber \\
&& \hspace{-10pt}=  {t}  +  \frac{1}{2}\left[\eta_{x,t} - \tilde{h}
 (x,0) +\sum_{y=1}^x y a_{y,t} + x \sum_{y=x+1}^La_{y,t}\right], \hspace{25pt} \elabel{eq_sol_H}
 \end{eqnarray}
 and with \bungledXR{\Eref{eq_def_S}}{Eq.~(3)} we obtain the expression of $S(t,M)$ in \bungledXR{\Esref{eq_SM}}{Eqs.~(4)} and \bungledXR{\eref{eq_xi1}}{(5)}.

\section{S3. Higher correlation functions of the interface}
\label{sect_app1}
As discussed in \cite{pruessner2012self}, the scaling in system size $L$ of the moments
$\langle s^n\rangle$ of the avalanche size of the one-dimensional boundary-driven
Oslo Model is $\langle s^n\rangle\propto L^{\sigma_n}$ with $\sigma_n=D(1+n-\tau)$
and $L$ the linear extent of the system. This scaling is a straightforward
consequence of the power-law distribution of the avalanche size,
\begin{equation}
P(s)=as^ {-\tau}\mathcal{G}\left(
\frac{s}{bL^D}\right)\ ,
\end{equation}
which applies to all $s$ beyond
a lower cutoff
$s_0$. Here,  $a$ and $b$ are metric factors and $\mathcal{G}$ is a scaling
function \cite{ChristensenETAL:1996, PhysRevE.94.042314, pruessner2012self}.
However, as we show here, the scaling form stated above is not enough
to characterise the moments of the multiple drive avalanche size $S(t,M)$.
The scaling argument $M/L^{D(\tau-1)}$ is essential for the scaling functions
considered in our Letter. The scale $L^{D(\tau-1)}$ is a correlation time
as introduced in \Eref{eq_corr}. We will demonstrate its necessity by firstly
attempting to derive the moments of the multiple
drive avalanches from the correlation function of the interface without
this additional scale, then introduce it and show how it reconciles the
interfacial correlation functions, the moments of $S$ and the established scaling
of $s$.

Firstly, we assume that the two-point, two-time connected correlation function
of $H$ has the usual (stationary) scaling form \cite{krug1997origins,
barabasi1995fractal}
\begin{equation}
\ave{H(x,t)H(x',t')}_c = a' L^{2\chi}\mathcal{C}_2\left(\frac{|t-t'|}{b'L^z},\frac{x}{L},\frac{x'}{L}\right),
\elabel{eq_2pointH_corr_func}
\end{equation}
where $\mathcal{C}_2$ is a scaling function, $a'$ and $b'$ are metric factors, $\chi$ is the roughness exponent and $z$ is the dynamical exponent which gives rise to the characteristic avalanche duration $\sim L^z$ in units of microscopic time steps.

\begin{widetext}
\begin{eqnarray}
\ave{S^2(t,M)} &=& \iint \ave{\big(H(x,t+M)-H(x,t)\big)\big(H(x',t+M)-H(x',t)\big)}\plaind x \plaind x'\elabel{eq_spat_integral0}\\
&=& 2\iint \big(\ave{H(x,t)H(x',t)}-\ave{H(x,t)H(x',t+M)}\big)\plaind x \plaind x' \elabel{eq_spat_integral}
\end{eqnarray}
\end{widetext}

Using \bungledXR{\Eref{eq_def_S}}{Eq.~(3)}, we derive that the scaling of the second moment $\ave{S^2(t,M)}$ can be calculated as in \Esref{eq_spat_integral0} and \eref{eq_spat_integral}
using uniformity at stationarity in the form
$\ave{H(x,t)}=\ave{H(x,t+M)}$ and
$\ave{H(x,t)H(x',t)}=\ave{H
(x,t+M)H(x',t+M)}$. Further using \Eref{eq_2pointH_corr_func} and the relation $D=\chi+d$ we obtain the scaling
$\ave{S^2(t,M)}=2 a'L^{2D} \left(\CCbar_2(0)-\CCbar_2\big(M/(b'L^z)\big)\right)$
with
$\CCbar_2(\tau)=\int\dint{y}\dint{y'} \mathcal{C}_2(\tau,y,y')$.
The difference between the two scaling functions, $\CCbar_2(0)-\CCbar_2\big(M/(b'L^z)\big)$, is, in fact, independent
of $M$. This is, firstly, because $b'L^z$ is the characteristic duration of
the avalanches. This is a \emph{microscopic} time and therefore a multiple
of $\delta t\ll 1$ and thus well separated from $M$ which is a \emph{macroscopic},
integer time.
On that timescale, the pile is no longer evolving and the interface is no longer relaxing, 
so that $\CCbar_2$ has converged in that argument. In other words, 
$\CCbar_2\big(M/(b'L^z)\big)$ should be considered as the limit $\tau\to\infty$
of $\CCbar_2(\tau)$ irrespective of $M$, provided $M$ is a positive integer.
Secondly, but closely related, $L^z$ governs the relaxation process, \ie the
avalanching, not the time $M$ that passes between drives. We conclude that
$\ave{S^2(t,M)}=2a' L^{2D} \big(\CCbar_2(0)-\lim_{\tau\to\infty}\CCbar_2(\tau)\big)\propto L^{2D}$.

This result is in contradiction with $\sigma_2=D(3-\tau)$ showing that the assumed scaling function in \Eref{eq_2pointH_corr_func} is, at least, incomplete.
To cure this contradiction, we need to consider the scaling of correlations on the macroscopic timescale. Its value $M^*\sim L^\chi$ is given by the characteristic number of avalanches needed to make interface configurations uncorrelated. In other words, $M^*$ is proportional to the characteristic scale of the vertical fluctuations of the interfaces $H(x,t)$ \cite{PhysRevLett.77.111}. Using $D=\chi+d$ and $\sigma_1=D(2-\tau)=d$ for $d=1$, we have $\chi = D(\tau-1)$. Hence, we amend the scaling function in \eref{eq_2pointH_corr_func} to obtain
\begin{eqnarray}
&&\ave{H(x,t)H(x',t')}_c \nonumber\\
&&= a' L^{2\chi}\mathcal{C}_2\left(\frac{|t-t'|}{b'L^z},\frac{|t-t'|}{c'L^{D(\tau-1)}},\frac{x}{L},\frac{x'}{L}\right),\elabel{eq_2pointH_corr_func_right}
\end{eqnarray}
with $c'$ a metric factor. The two timescales $L^z$ and $L^{D(\tau-1)}$ are
proportional to the microscopic and the macroscopic timescale respectively.
As we consider $|t-t'|=M$ in the following, which is large compared to the
microscopic timescale $b'L^z$ as discussed above, we will drop this argument
in what follows.

After performing the spatial integral \eref{eq_spat_integral}, the scaling
of the second moment is 
\begin{equation}\elabel{correct_S2}
\ave{S^2(t,M)}= a'L^{2D}\left(\tilde{\mathcal{C}}_2
(0)-\tilde{\mathcal{C}}_2\left(M/\left(c'L^{D(\tau-1)}\right)\right)\right)\ ,
\end{equation}
where $\tilde{\mathcal{C}}_2\left(\tau'\right)=\int\dint{y}\dint{y'} \mathcal{C}_2(\tau,\tau',y,y')$.
\Eref{correct_S2} 
 can be reconciled with the scaling of $
\ave{S^2(t,1)}=\ave{s^2}\propto L^{D(3-\tau)}$ only if $\tilde{\mathcal{C}}_2\left(0\right)-\tilde{\mathcal{C}}_2\left(\tau'\right)$
is linear for small arguments $\tau'=1/(c'L^{D(\tau-1)})$. Therefore, the scaling
of the second moment 
is $\ave{S^2(t,M)}\propto ML^{D(3-\tau)}\hat{\mathcal{C}}_2\left(M/c'L^{D
(\tau-1)}\right)$ with $\hat{\mathcal{C}}_2\left(\tau'\right)=\left(\tilde{\mathcal{C}}_2\left(0\right)-\tilde{\mathcal{C}}_2\left(\tau'\right)\right)/\tau'$ convergent in small arguments
$\tau'$.

Indeed, applying this argument to the $n$th moment, we obtain the scaling
$\ave{S^n(t,M)}\propto ML^{D(1+n-\tau)}\hat{\mathcal{C}}_n\left(M/c'L^{D(\tau-1)}\right)$,
which is consistent with the established exponents $\sigma_n$, see \bungledXR
{}{Fig.~3}. Moreover, the numerics are consistent with that $\ave{S^n(t,M)}\propto M$
for small $M$ and all $n$, as shown in \bungledXR{}{Fig.~3}.

 \section{S4. Uniform drive in the bulk}
 \label{sect_app_bulk}
 
 \begin{figure}
 \centering
\subfigure[\hspace{230pt}]{ \label{fig_oobulka} \includegraphics[width=0.48\textwidth]{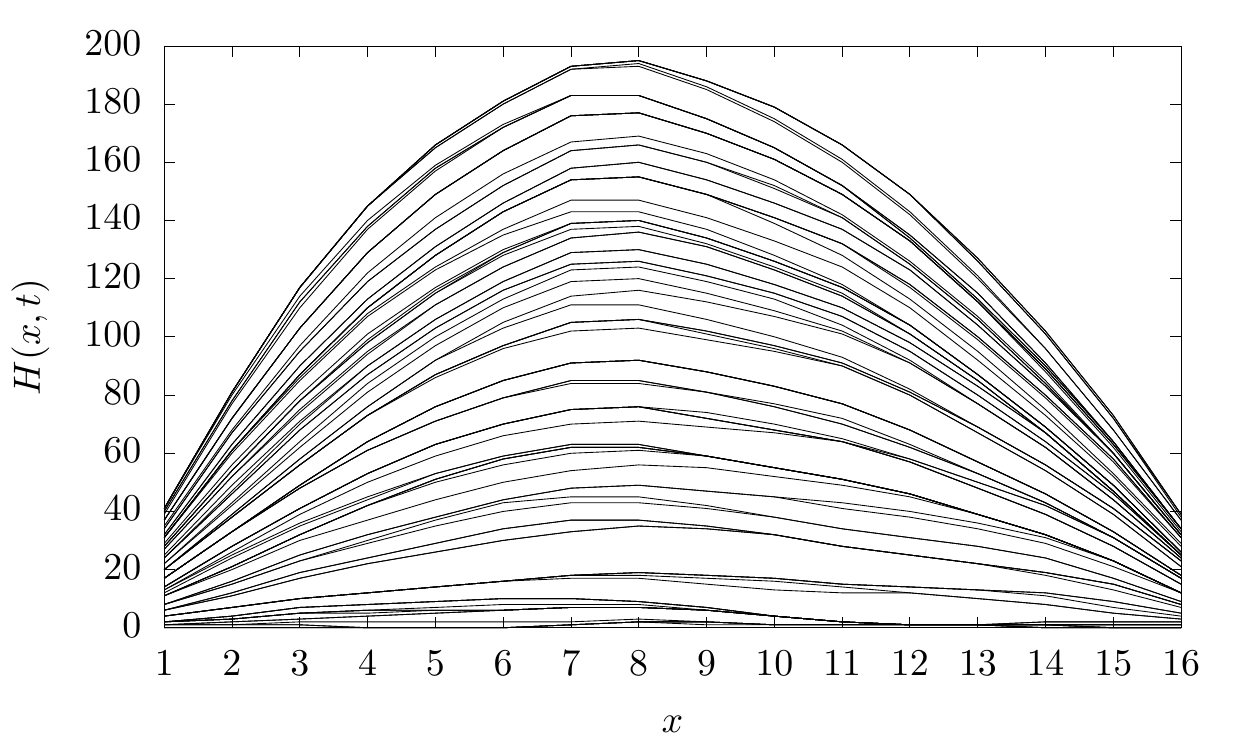}}
\subfigure[\hspace{230pt}]{ \label{fig_oobulkb} \includegraphics[width=0.48\textwidth]{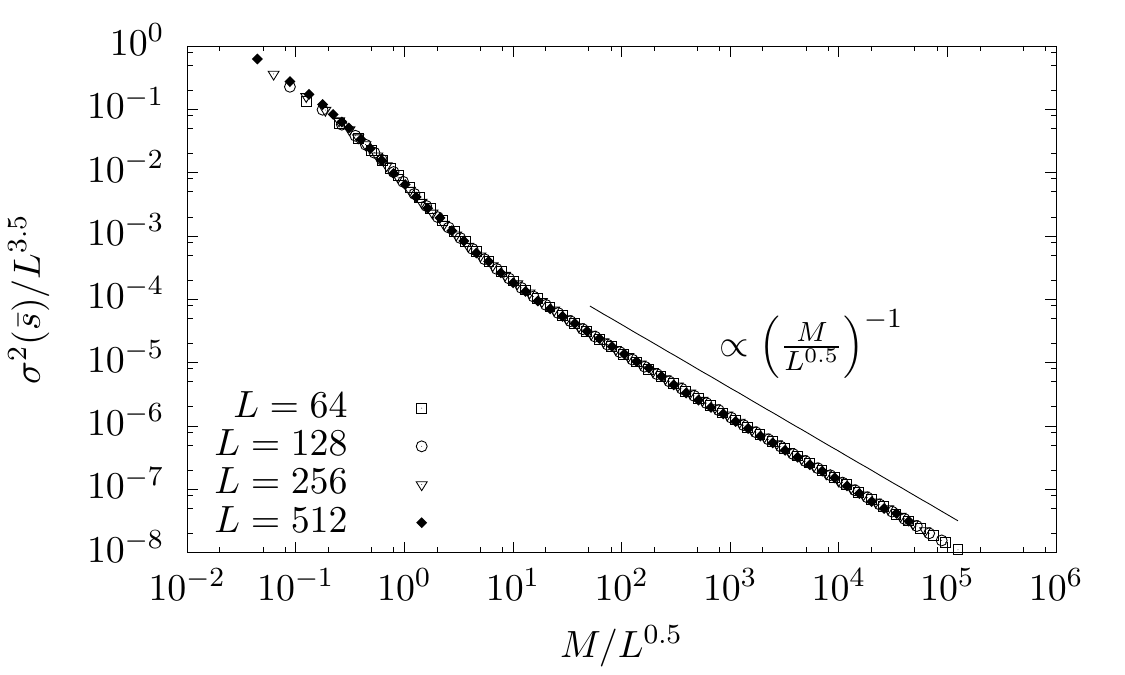}}
\caption{ \label{fig_oobulk} One-dimensional bulk driven Oslo Model with open boundary conditions. \subref{fig_oobulka} Typical interface configurations of topplings $H(x,t)$ in a system of size $L=16$ with $t\in[0,90]$ and initial condition $z_x=1$ for all $x$; \subref{fig_oobulkb} The scaling of the variance $\sigma^2(\bar{s})$ is asymptotically proportional to $M^{-1}L^4$ (measurements have been taken starting from recurrent configurations).}
\end{figure}

 In this section we derive the equations of motion of $h(x,t)$ and $H(x,t)$
 and the multiple drive avalanche size $S(t,M)$ for the one-dimensional bulk driven Oslo Model on a lattice with open boundaries \cite{PhysRevE.78.041102, pruessner2012self}, along the lines of Sec.~S2.
 The boundaries are now both open, \latin{i.e.}~$h(0,t)=h(L+1,t)=0$ for all $t$ and the equation
 of motion has a source term given by $E(x,t) = \sum_{j=1}^t\theta
 (t-j)\delta_{x,x_j}$, where $\theta(t)$ is the Heaviside step function, $\delta_
 {x,y}$ is the Kronecker delta and $x_j\in\{1,\ldots,L\}$ accounts for the
 randomly chosen site for driving the pile at each time step $j=1,\ldots,t$.
 Hence, the source function $E(x,t)$ counts the number of external grains
 that have been dropped by the external drive at $x$ up to time $t$. The equation
 of motion now reads
\begin{eqnarray}
&&{h}({x,t+\delta t})-{h}({x,t})  \nonumber\\
&&=E(x,t)+\frac{1}{2}\big[{h}({x-1,t}) - 2{h}({x,t}) +{h}({x+1,t})\big.\nonumber\\
&& \big.\hspace{10pt}+ \eta_{x-1,t} + \eta_{x+1,t}-\eta_{x-1,0} - \eta_{x+1,0}\big]. \elabel{eq_SEOM_bulk}
\end{eqnarray}
At quiescence, \Eref{eq_SEOM_bulk} with boundary conditions as stated above has the solution
\begin{eqnarray}
\hspace{-10pt}h(x,t)  &=& \left(\frac{L+1-x}{L+1}\right)\sum_{y=1}^{x-1}yb_{y,t}\nonumber\\
&&\hspace{-10pt}+\frac{x}{L+1}\sum_
{y=x}^L(L+1-y)b_{y,t}-\hat{h}(x,0), \elabel{eq_h_bulk}
\end{eqnarray}
where 
\begin{equation}
b_{x,t}=-\Delta h(x,t)_{|_{t\in\mathbb{Z}}}=2E(x,t)+\eta_{x-1,t}+\eta_{x+1,t},
\end{equation}
and
\begin{eqnarray}
\hat{h}(x,0)&=&\left(\frac{L+1-x}{L+1}\right)\sum_{y=1}^{x-1}yb_{y,0}\nonumber\\
&&+\frac{x}{L+1}\sum_{y=x}^L(L+1-y)b_{y,0}\ ,
\end{eqnarray}
accounting for the initialisation.
From \Esref{eq_Hh} and \eref{eq_h_bulk} we can write $H(x,t)$ in closed form, see Fig.~S\ref{fig_oobulka}, and further using \bungledXR{\Eref{eq_def_S}}{Eq.~(3)} we derive the multiple drive avalanche size (\cf \bungledXR{}{Eq.~(4)})
\begin{equation}
S(t,M) =\Lambda(t,M)+\xi'(t,M),
\elabel{eq_S_bulk}
\end{equation}
where the random variable
\begin{equation}
\Lambda(t,M) = \frac{1}{2} \sum_{x=1}^L x(L+1-x)\big(E({x,t+{M}})-E({x,t}) \big)
\elabel{eq_Lambda}
\end{equation}
is due to the drive and
\begin{eqnarray}
&&\hspace{-15pt}\xi'(t,M) = \frac{1}{2} \sum_{x=1}^L \Big[\eta_{x,t+{M}}-\eta_{x,t}+\frac{x}{2}(L+1-x) \Big.\nonumber\\
&&\Big.\times\left( \eta_{x-1,t+{M}}+ \eta_{x+1,t+{M}} - \eta_{x-1,t}-\eta_{x+1,t}\right)\Big]\hspace{15pt}\elabel{eq_xip}
\end{eqnarray}
is the new noise term. The difference between $\xi'$ in \eref{eq_xip} and the noise $\xi$ in \bungledXR{\Eref{eq_xi1}}{(5)} is exclusively due to the different boundary conditions applied to the interfaces. It turns out that $\xi'$ is also a finite sum of bounded random variables and thus $\xi'$ is bounded itself. Therefore, $\xi'$ is asymptotically independent of $t$ and $M$ implying that  $\sigma^2\left(\xi'\right)$ tends to a constant as $M\to\infty$.

From \Eref{eq_S_bulk} we have that $\sigma^2(S(t,M))=\sigma^2\left(\Lambda\right) +\sigma^2\left(\xi'\right) +2\text{Cov}(\Lambda,\xi')$. The variance of $\Lambda(t,M)$ can be calculated using the fact that the vector of random variables $\mathbf{E}(t+M)-\mathbf{E}(t)$ has a multinomial distribution independent of $t$ with probabilities $p_x=1/L$ for all $x$ and $M$ trials, which gives
\begin{equation}
\sigma^2\big(\Lambda(t,M)\big) =\frac{1}{720}M\left(L^4-5L^2+4\right).
\elabel{eq_simga_L}
\end{equation}
Owing to the form of $\Lambda$ and $\xi'$, it follows that $\text{Cov}(\Lambda,\xi')$ is subleading in $M$ compared to $\sigma^2(\Lambda)$ as $M\to\infty$.
As $\sigma^2(\xi')$ converges in $M$, the variance $\sigma^2(S(t,M))\propto
ML^4$ so that $\sigma^2(\bar{s})\propto M^{-1}L^4$, implying that the avalanche
size of the uniformly driven Oslo Model does not display hyperuniformity,
but rather Poisson-like behaviour.

Our numerical results, Fig.~S\ref{fig_oobulkb}, show that the variance follows the asymptotic scaling form
$\sigma^2(\bar{s}) = L^{3.5}\tilde{\mathcal{D}}\left({M}/{L^{0.5}}\right)$,
with $\tilde{\mathcal{D}}(x)\propto x^{-1}$ as $x\to\infty$. That is, asymptotically, $\sigma^2(\bar{s})\propto M^{-1}L^4$. In fact, at large $M$ the scaling is solely due to \Eref{eq_simga_L} as $\sigma^2(\bar{s})$ is dominated entirely by $\sigma^2(\Lambda)/M^2$. The interfaces detach dominantly due to the noise in the driving rather than the relaxation dynamics of the interface in response to the boundary driving.

In general, if the driving position $x$ is fixed, then $\Lambda(t,M)=\frac{1}{2}x(L+1-x)M$ becomes deterministic so that $\sigma^2(\Lambda)=0$ and the results for the boundary-driven Oslo Model generalise to the Oslo Model driven at a fixed (bulk) site. In this case, the avalanche size is hyperuniform on long timescales regardless of the driving position $x$ and the boundary conditions. In particular, if the drive is located at the boundary $x=1$ then $\Lambda(t,M)=\frac{1}{2}LM$, where the difference in the factor $1/2$ in comparison to \bungledXR{\Eref{eq_SM}}{Eq.~(4)} is due to the different boundary conditions.

\section{S5. Two-dimensional Oslo Model}
\label{sect_app_2d}
We have also studied the fluctuations in the avalanche size in the two following settings on a square lattice: (a) two open and two closed boundaries with the drive located at the intersection of two open boundaries (corner drive) and (b) four open boundaries and uniformly distributed drive on the lattice (bulk drive). These two settings resemble the one-dimensional cases studied in this Letter.

\begin{figure}
\centering
\subfigure[\hspace{230pt}]{ \label{fig_2d_a} \includegraphics[width=0.47\textwidth]{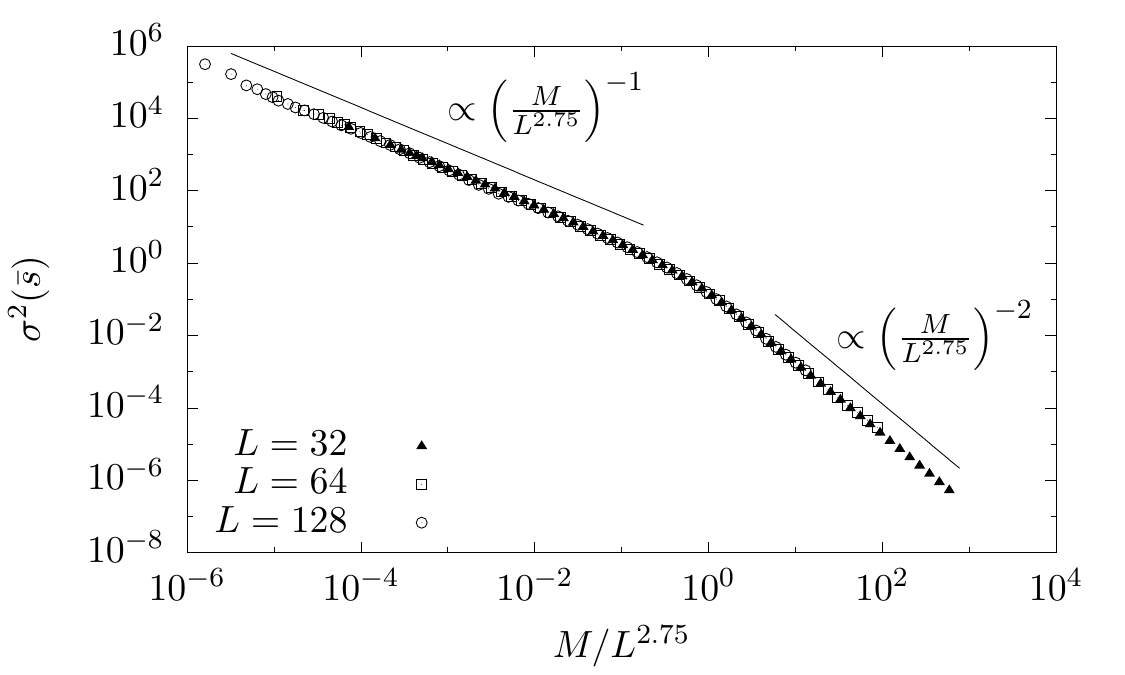} }
\subfigure[\hspace{230pt}]{ \label{fig_2d_b} \includegraphics[width=0.47\textwidth]{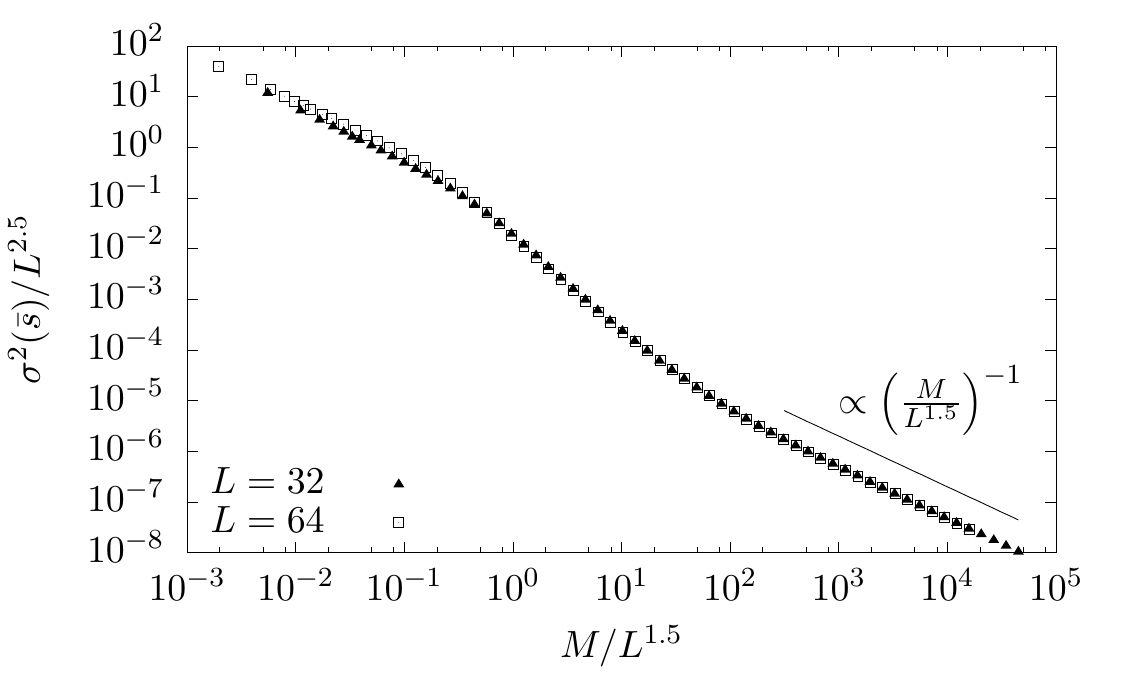} }
\caption{ \label{fig_2d} Scaling of $\sigma^2(\bar{s})$ in the two-dimensional Oslo Model with \subref{fig_2d_a} two open and two closed boundaries and the drive located at the intersection of the two open boundaries (corner drive, \cf \bungledXR{\Figref{Dx}}{Fig.~2}) and \subref{fig_2d_b} four open boundaries with drive uniformly distributed in the bulk (see \Figref{oobulkb}). In all cases, the pile is initialised in a recurrent configuration. Similarly to the one-dimensional cases, in \subref{fig_2d_a} the fluctuations $\sigma^2(\bar{s})$ have two scaling regimes, displaying hyperuniformity for $M\gg L^{2.75}$, and in \subref{fig_2d_b} $\sigma^2(\bar{s})$ is asymptotically proportional to $M^{-1}L^4$ indicating the absence of hyperuniformity.}
\end{figure}

Our numerical results for setting (a) are shown in Fig.~S\ref{fig_2d_a}. The variance has the scaling form
\begin{equation}
\sigma^2(\bar{s}) = \mathcal{D}\left(\frac{M}{L^{2.75}}\right),
\end{equation}
with $\mathcal{D}(x)\sim x^{-1}$ for $x\ll1$ and $\mathcal{D}(x)\sim x^{-2}$ for $x\gg1$. The exponent $2.75$ is in fact the avalanche dimension
$D=2.75\pm0.01$ \cite{pruessner2012self, PhysRevE.78.041102, lubeck2000moment,
paczuski1996avalanche}, which enters into the characteristic number of driving steps, $M^*\sim L^{D(\tau-1)}$. Since the average avalanche size diverges only logarithmically \cite{pruessner2012self}, $\sigma_1=D(2-\tau)=0$, in the case of corner drive with boundary conditions as described above, it follows that $\tau=2$, hence $D(\tau-1)=D$ rather than $D(\tau -1) = \chi$ as in the one-dimensional case. Further, the second moment of $S$ scales like $ML^{D(3-\tau)}=ML^D = M^2\left(M/L^D\right)^{-1}$, so that
\begin{equation}
\sigma^2(\bar{s}) \propto \frac{1}{M^2}\left\langle S^2(t,M)\right\rangle = \mathcal{D}\left(\frac{M}{L^{D}}\right).
\end{equation}

In setting (b), Fig.~S\ref{fig_2d_b}, we find numerically that the variance has the scaling form
$\sigma^2(\bar{s}) = L^{2.5}\tilde{\mathcal{D}}\left({M}/{L^{1.5}}\right)$,
with $\tilde{\mathcal{D}}(x)\sim x^{-1}$ as $x\to\infty$,
following the same scaling as the one-dimensional bulk driven Oslo Model. We obtain $\sigma^2(\bar{s})\propto M^{-1}L^4$ in large $M$, as, again, fluctuations of $\bar{s}$ are dominated by the driving rather than the interface roughness.

\end{document}